\documentclass{phb-proc4-auth}

\usepackage{graphicx}
\usepackage{amssymb}
\usepackage{amsmath}

\newcommand{\Journal}[4]{#1 \textbf{#2}, #3 (#4)}
\newcommand{\eps}{\varepsilon}
\newcommand{\mb}[1]{\mathbf{#1}}

\newcommand{\yt}{(Y$_{1-x}$Th$_x$)$_2$C$_{3-y}$}
\newcommand{\la}{(La$_{1-x}$Th$_x$)$_2$C$_{3-y}$}

\newcommand{\be}{\begin{equation}}
\newcommand{\ee}{\end{equation}}

\begin{document}
\begin{frontmatter}


\journal{SCES '04}


\title{Are Th-doped Y$_2$C$_3$ and La$_2$C$_3$ two-band superconductors?}

%

\author{Ivan A. Sergienko\corauthref{adr}\thanksref{nserc}}

 
\address{Department of Physics \& Physical Oceanography, 
Memorial University of Newfoundland, St. John's, NL A1B 3X7, Canada}

%

\thanks[nserc]{This work was supported by NSERC of Canada.}

%

\corauth[adr]{Present address: Condensed Matter Science Division,
Oak Ridge National Laboratory, Oak Ridge, TN 37831, USA.\\ 
Electronic address: sergienko@ornl.gov}


\date{15 June 2004}

\begin{abstract}
The superconducting sesquicarbides R$_2$C$_3$ have noncentrosymmetric 
point group symmetry $T_d$. Spin-orbit coupling lifts the spin degeneracy of 
electronic bands in the most of the Brillouin zone. Nevertheless, due to high 
symmetry, there are a few directions along which the Fermi surfaces must touch. 
This leads to two-band effects in the superconducting state.
\end{abstract}

%



\end{frontmatter}

%

The recent discovery of heavy-fermion superconductivity in the noncentrosymmetric 
tetragonal compound CePt$_3$Si~\cite{Bauer04} has raised a new challenge concerning 
the 
nature of superconducting states~\cite{Frigeri04,Sergienko04}. This is due to 
the lifting of spin degeneracy of electronic bands in noncentrosymmetric crystals
by strong spin-orbit coupling 
(SOC). This note is devoted to two series of compounds  with so-called
sesquicarbide structure, \yt\ and \la, discovered more than three decades 
ago~\cite{Krupka,Giorgi}. The crystals appeared to be superconducting with $T_c$ up
to 17 K. The sesquicarbides have space group $I\bar43d$~\cite{Amano04}, which 
belongs to the tetrahedral crystallographic class $T_d$. 
Centres of symmetry are therefore absent from the 
structure. Th doping ($x$ can reach 0.9) introduces two features which make the
sesquicarbides similar to CePt$_3$Si, namely Th has the same number of 
$f$-electrons in the outer shell as Ce, and also one can expect strong SOC effects.
Unfortunately, the physical properties of the sesquicabides have not been studied
systematically so far. 

Here I predict that superconductivity in Th-doped sesquicarbides
should involve at least two
bands with possibly different magnitudes of the superconducting gap. If the Fermi 
surfaces corresponding to different bands split by strong SOC  are separated, a 
one-band treatment may be sufficient~\cite{Sergienko04}. 
For \yt\ and \la, there are directions in the Brillouin zone where the SOC split
Fermi surfaces must touch. This is clear from the following group theoretical
argument. The electronic states are described by the Bloch spinors
\be\label{spinors}
\Psi^\pm_{\mb k}(\mb r) = U^\pm_\mb k(\mb r) e^{i \mb k \mb r},
\quad
U^\pm_\mb k(\mb r) = \frac{1}{\sqrt{2V}} \left(
\begin{array}{c}u_{\mb k, \uparrow}\\
u^\pm_{\mb k, \downarrow} \end{array} \right),
\ee 
where $\mb k$ is quasimomentum, $V$ is the sample volume and indexes $\pm$ 
refer to the two 
SOC split bands~\cite{Sergienko04}. Just as in the case of zero
SOC, the functions $U^\pm_\mb k(\mb r)$ span irreducible
representations of the point group which leaves the vector $\mb k$
invariant. However, in the present case the representations are double 
valued since $U^\pm_\mb k(\mb r)$ describe spin-$\frac 1 2$ particles.
For $T_d$ symmetry, the vectors $\mb k_1 = (k_x,0,0), (0,k_y,0), (0,0,k_z)$ are 
invariant under the group $C_{2v} \subset T_d$. 
This group has only one double valued irreducible representation, which is
two-dimensional.\footnote{Note that one-dimensional representations should 
not be combined with their complex conjugate since time reversal changes the 
sign of $\mb k$.} 
Hence, there must be two linearly independent solutions of the Schr\"odinger 
equation corresponding to each allowed energy value $\eps_{\mb k_1}$. 

For the sake of simplicity, I further proceed with addressing the effects
of SOC by the perturbative approach proposed in~\cite{Frigeri04}.
The single particle Hamiltonian is approximated by
\be\label{Ham1}
H_1 = \sum_{\mb k, s} \eps^0_{\mb k} c^\dagger_{\mb k s} c_{\mb k s} +
\alpha \sum_{\mb k, s, s'} \boldsymbol g_{\mb k} \, \boldsymbol \sigma_{s s'}
c^\dagger_{\mb k s} c_{\mb k s'},
\ee
where $\alpha>0$ is constant, 
$\boldsymbol \sigma =(\sigma_x,\sigma_y,\sigma_z)$ are the Pauli matrices and 
the vector $\boldsymbol g_{\mb k} = [g_x(\mb k), g_y(\mb k), g_z(\mb k)]$ is 
chosen such that 
$\boldsymbol g_{\mb k} \, \boldsymbol \sigma$ reduces the symmetry of
$H_1$ from $G\otimes I$ to $G$, the actual point group. Here $I$ is inversion.
The Hamiltonian~(\ref{Ham1}) is diagonalized by the two spinors~(\ref{spinors}) 
with 
\be\label{spinor1}
u^\pm_{\mb k, \downarrow}/u_{\mb k, \uparrow} = 
(\pm |\boldsymbol g_{\mb k}| - g_z)/(g_x-ig_y),
\ee 
corresponding to the eigenvalues 
\be\label{bands}
\eps^\pm_{\mb k} = \eps^0_{\mb k} 
\pm \alpha |\boldsymbol g_\mb k|.
\ee
It is a somewhat lengthy but straightforward algebraic exercise to show that 
\be\label{spin}
\mb s^\pm_{\mb k} \equiv \frac{\hbar}{2} \langle \Psi^\pm_{\mb k}| \boldsymbol 
\sigma |\Psi_{\mb k}^\pm\rangle = \pm \frac{\hbar}{2} \frac{\boldsymbol g_{\mb k}}
{|\boldsymbol g_{\mb k}|}.
\ee
Hence, $\boldsymbol g_{\mb k}$ defines the direction of spin 
$\mb s^\pm_{\mb k}$ carried by particles in both bands.

\begin{figure}
\includegraphics[width=\columnwidth]{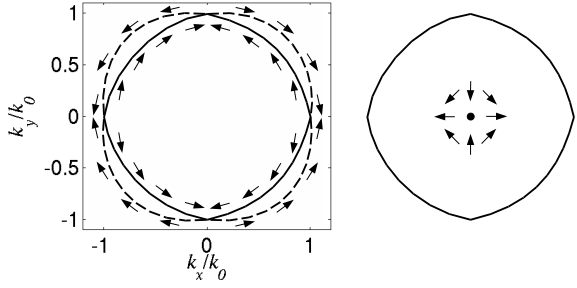}
\caption{\label{fig1}\emph{Left}. Cross section of the Fermi surfaces by the plane 
$k_z=0$. Solid line corresponds to the '+' band, dashed line -- the '-' band.
Arrows denote the direction of spin. \emph{Right}. Spins in the '+' band
around a node of $\boldsymbol g_\mb k$. }
\end{figure}

For $G=T_d$, $G\otimes I = O_h$, and $\boldsymbol g_{\mb k} \, 
\boldsymbol \sigma$ transforms according to the representation $A_{2u}$
of $O_h$. Namely,
\be\label{gvec}
\boldsymbol g_{\mb k} = [k_x (k_z^2-k_y^2), k_y(k_x^2-k_z^2), k_z (k_y^2-k_x^2)]
\ee
It follows from~(\ref{spin}) that the points $\boldsymbol g_\mb k=0$, where the 
Fermi surfaces touch, are singular, \emph{i. e.} the direction of spin is not 
defined. This is essentially because the spin quantization axis can be chosen 
arbitrary for spin degenerate states. A cross section of the Fermi surfaces 
corresponding to~(\ref{gvec}) is shown in Fig.~\ref{fig1} together with the 
spin structure. It is assumed that the ``unperturbed'' Fermi surface is spherical,
$\eps_\mb k^0 = \hbar^2 k_0^2/2m$. 

Finally, I consider the pairing interaction term $H_2$ in the Hamiltonian, which 
leads to the formation of the superconducting state. I assume that before SOC has 
been turned on, the electrons were paired in a singlet superconducting state, with 
the order parameter $\psi(\mb k)$. Within the weak coupling 
approach, 
\be\label{ham2}
H_2=\frac 1 2 \sum_{\mb k} \psi(\mb k) (c^\dagger_{\mb k \uparrow} 
c^\dagger_{-\mb k \downarrow} - c^\dagger_{\mb k \downarrow} 
c^\dagger_{-\mb k \uparrow}) + \text{h.c.}
\ee
If by some reason it is desirable, one can start with a triplet pairing state.
As far as the symmetry of the superconducting state is concerned, the result will 
be essentially the same~\cite{Sergienko04}.

Equation~(\ref{ham2}) is valid for the degenerate states at the nodes of 
$\boldsymbol g_\mb k$. However, for the rest of the Brillouin zone $H_2$ should be 
expressed in terms of the new fermion operators $c^\dagger_{\mb k +}$, 
$c^\dagger_{\mb k -}$ corresponding to the bands~(\ref{bands}). 
I introduce the spherical coordinate system $k_x = \cos \theta$, 
$k_z = \sin \theta \cos \phi$, $k_y = \sin \theta \sin \phi$ and consider the 
states in the vicinity of $\mb k_1=(k_x,0,0)$.
Then using~(\ref{spinor1}), I obtain
\be
\begin{array}{rcl}
c^\dagger_{\mb k \uparrow} &=& c^\dagger_{\mb k +} \sin 
(\phi/2) + 
c^\dagger_{\mb k -} \cos 
(\phi/2), \\
c^\dagger_{\mb k \downarrow} &=&-i c^\dagger_{\mb k +} \cos 
(\phi/2) + 
i c^\dagger_{\mb k -} \sin 
(\phi/2).
\end{array}
\ee
Hence, the pairing interaction is   
\be\label{ham2new}
H_2 = \frac i 2 \sum_{\mb k} \psi(\mb k) (c^\dagger_{\mb k +} c^\dagger_{-\mb k +} 
+ c^\dagger_{\mb k -} c^\dagger_{-\mb k -}) + \text{h.c.}
\ee
The factor $i$ is the additional phase factor $t(\mb k)$ which is acquired by the
gap function in the superconductors with SOC split bands~\cite{Sergienko04}.
Generally, $t(\mb k)$ is an odd function, which follows from the anticommutation
of the fermion operators. Equation~(\ref{ham2new}) is written for $\mb k$ with
$\theta\approx 0$, and one cannot replace $\mb k$ with $-\mb k$ in the sum. 
For $\theta\approx \pi$, the right hand side in~(\ref{ham2new})
changes its sign, and thus the oddness of $t(\mb k)$ is restored.

Hence, even though there is no interband coupling terms in~(\ref{ham2new}),
the pairing in the two bands is governed by the same order parameter
$\psi({\mb k})$, that is both gaps open at the same $T_c$. However, since 
$\psi(\mb k)$ at least depends on the absolute value of $\mb k$, the amplitudes of
the gaps away from the nodes of $\boldsymbol g_\mb k$ may be quite different, 
similar to the situation in MgB$_2$~\cite{mgb}.

In conclusion, it is shown that the Fermi surfaces of \yt\ and \la\ that are split 
by SOC must touch along some directions in the Brillouin zone due to their high 
crystallographic symmetry. This implies that the minimal model should at least 
include two bands with the superconducting gaps opening at the same $T_c$.
I hope that this note will stimulate interest towards a more active investigation 
of the non-trivial superconducting properties of Th-doped sesquicarbides.

I thank S.H. Curnoe and D.J. Singh for useful discussions.

%



\begin{thebibliography}{0}

\bibitem{Bauer04} E. Bauer \emph{et al.}, 
\Journal{Phys. Rev. Lett.}{92}{027003}{2004}.

\bibitem{Frigeri04} P.A. Frigeri \emph{et al.}, \Journal{Phys. Rev. Lett.}{92}
{097001}{2004}.

\bibitem{Sergienko04} I.A. Sergienko and S.H. Curnoe, cond-mat/0406003.

\bibitem{Krupka} M.C. Krupka \emph{et al.}, \Journal{J. Less-Common Met.}{19}{113}
{1969}; \Journal{\emph{ibid}}{17}{91}{1969}.

\bibitem{Giorgi} A.L. Giorgi \emph{et al.}, \Journal{J. Less-Common Met.}{17}{121}
{1969}; \Journal{\emph{ibid}}{22}{131}{1970}.

\bibitem{Amano04} G. Amano \emph{et al.}, \Journal{J. Phys. Soc. Jpn.}{73}{530}
{2004}.

\bibitem{mgb} M. Iavarone \emph{et al.}, \Journal{Phys. Rev. Lett.}{89}
{187002}{2002}.

\end{thebibliography}
\end{document}